\documentstyle[12pt,aaspp4]{article}

\newcommand{\gtilde}
 {~ \raisebox{-1ex}{$\stackrel{\textstyle >}{\sim}$} ~}
\newcommand{\ltilde}
 {~ \raisebox{-1ex}{$\stackrel{\textstyle <}{\sim}$} ~}

\begin{document}

\title{Very Strong TeV Emission as Gamma-Ray Burst Afterglows}

\author{Tomonori Totani}
\affil{Department of Physics, School of Science,
The University of Tokyo, Tokyo 113-0033,
Japan \\ E-mail: totani@utaphp2.phys.s.u-tokyo.ac.jp}

\begin{abstract}
Gamma-ray bursts (GRBs) and following afterglows are considered to
be produced by dissipation of kinetic energy of a relativistic
fireball and radiation process is widely believed as
synchrotron radiation or inverse Compton scattering of electrons. 
We argue that the transfer of kinetic energy of ejecta into electrons
may be inefficient process
and hence the total energy released by a GRB event
is much larger than that emitted in soft gamma-rays, by a factor
of $\sim (m_p/m_e)$. We show that,
in this case, very strong emission of TeV gamma-rays is possible
due to synchrotron radiation 
of protons accelerated up to $\sim 10^{21}$ eV,
which are trapped in the magnetic field of afterglow shock and 
radiate their energy on an observational time scale of $\sim$ day.
This suggests a possibility that GRBs are most energetic in TeV range and 
such TeV gamma-rays may be detectable from GRBs even at 
cosmological distances,
i.e., $z \sim 1$, by currently working ground-based telescopes.
Furthermore, this model gives a  quantitative explanation for 
the famous long-duration GeV photons detected from
GRB940217. If TeV gamma-ray emission which
is much more energetic than GRB photons is detected, 
it provides a strong evidence for acceleration 
of protons up to $\sim 10^{21}$ eV.
\end{abstract}


\section{Introduction}
Gamma-ray bursts (GRBs) are widely believed as
dissipation of kinetic energy of relativistic motion
produced by an expanding fireball
with a Lorentz factor of $\sim 10^2$--$10^3$
(see e.g., Piran 1994 for a review). The recently discovered afterglows
following GRBs are also considered as similar phenomena,
which are dissipation
in the external shock generated by the collision with interstellar matter
(Paczy\'nski \& Rhoads 1993; Katz 1994;
M\'esz\'aros \& Rees 1997; Vietri 1997a).
The cosmological origin of GRBs is now almost confirmed by the discovery
of metal absorption lines at $z$= 0.835 for the optical afterglow
of GRB970508 (Metzger et al. 1997), and some of observations for
X-ray, optical, and radio afterglows are in rough agreement with
the prediction of the cosmological fireball model
(Wijers, Rees, M\'esz\'aros 1997; Waxman 1997a, b; Vietri 1997b). 
However, there is large variation in the afterglow response of GRBs
(e.g., Groot et al. 1998), and it is not yet clear whether the 
simple afterglow model is applicable for all of GRBs.

There are two important, but highly uncertain parameters in such theoretical
models of GRBs and afterglows: the degree of equipartition between
the internal energy of shock heated matter and magnetic fields
($\xi_B$) and between protons and electrons
($\xi_e$). In most of publications which calculated model predictions
of GRBs or afterglows, these two parameters are assumed to be
of order unity, and the radiation process is considered as
electron synchrotron (or inverse Compton scattering). In this case
the efficiency of energy release in GRBs or afterglows compared to
the total energy of a GRB event ($E$) is
of order unity. However, currently there is no clear evidence for efficient
energy transfer into electrons and magnetic fileds, although
some of observational data are consistent with $\xi_e \sim 1$ (Waxman 1997a).
If the energy transfer from protons into electrons is inefficient, 
energy stored in electrons is only a fraction of $\xi_e \sim (m_e/m_p)$ of the 
total fireball energy and hence about 2000 times larger energy
must be released as kinetic energy of relativistic ejecta
than the observed energy emitted as GRB photons. The GRB photon energy is
$\sim 10^{51}$ erg
if the radiation is isotropic and the redshift of most distant
GRBs, $z_{\max} \sim$ 1.
Then the total energy $E$ may be uncomfortably large because
most of GRB models are based on gravitational collapses
of massive stars in which available energy is $\sim 3 \times 
10^{53}$ erg and most of
this energy will be lost as neutrinos. However, the theoretical estimate of
merger rate of binary neutron stars is about $10^{2-3}$ times higher than
the observed GRB rate (Lipunov et al. 
1995; Totani 1997,1998), which suggests that GRB is strongly beamed 
if GRBs are associated to merger of binary neutron stars
(Blinnikov et al. 1984). If GRBs are
actually beamed with such a strong beaming factor, the above constraint
of energy budget becomes much weaker. Much more energetic
models of GRBs have also been proposed such as the microquasar model,
in which total energy of $\sim 10^{54}$ erg can be supplied to a fireball
(Paczy\'nski 1998). 

In this letter we argue that time scale 
of energy transfer into electrons by the Coulomb interactions
is much larger than the expansion time of external shock, while
magnetic field may achieve the equipartition with protons in
the shock heated matter. We then show that, as a consequence of this
scenario, a very strong TeV emission is expected during a few days
after GRBs by synchrotron radiation of $10^{20}$ eV protons and it may be
detectable by current ground-based
telescopes even from a GRB at cosmological distances, in spite of
significant attenuation due to $e^\pm$ creation with intergalactic 
infrared photons. 
Synchrotron emission of protons of $\sim 10^{20}$ eV from GRBs was first
considered by Vietri (1997c), and B\"ottcher \& Dermer (1998)
extended the analysis to emission from afterglows. Both papers
concluded that TeV gamma-rays are detectable only for nearby GRBs
($z \ltilde 0.1$), assuming that total fireball energy is of the same
order with that of GRB photons. The natural units with
$c = \hbar = 1$ is used in this letter.

\section{Efficiency of Energy Transfer into Electrons and Magnetic Fileds}
The evolution of external shock is described by 
$bE = 16 \pi n m_p r^3 \gamma^2 /17$, where $E$ is the total energy
released in an opening angle of $\Delta \Omega$, $b \ (= 4\pi/\Delta \Omega)$
a beaming factor, and $\gamma$ the Lorentz factor of the shock heated
matter (Blandford \& McKee 1976). 
The location of the shock, $r$, is measured in the laboratory
frame and $n$ is the (unshocked) interstellar matter density.
Initially the kinetic energy stored in electrons
is only a fraction of $(m_e/m_p)$ of the total energy, and
much greater energy of protons must be efficiently transferred into electrons
by some interactions in the shock heated matter in order to 
achieve energy equipartition between electrons and protons. However, 
relative importance of the Coulomb interaction becomes
smaller with increasing energy of particles, and the particle energy
in relativistic shocks is much greater than that in non-relativistic shocks
such as supernova remnants. The time scale of energy transfer
in relativistic plasma is difficult to estimate accurately, but
a rough estimate is given by $\tau_{ep} \sim (n' \sigma_t)^{-1}$, where 
$n' = 4 \gamma n$ is the proton number density
of the shocked matter measured in the shocked-shell frame and 
$\sigma_t = 4 \pi L_e (e^2/m_e\gamma)^2$ is the transport cross section
for electron-proton collisions. The Coulomb logarithm
is given by $L_e = \ln (a m_e \gamma)$, where $a = 
(m_e \gamma/4\pi n'e^2)^{1/2}$ 
is the Debye length (e.g., Lifshitz \& Pitaevskii 1981).
This time scale should be compared to
the expansion time measured in the shell frame, $r/\gamma$,
and we find $\tau_{ep}/(r/\gamma) = 1.1 \times 10^4 b^{-1/3}
E_{51}^{-1/3} n_1^{-2/3} \gamma^{8/3}$, where $E = 10^{51} E_{51}$ erg
and $n = n_1 \ \rm cm^{-3}$.
Hence energy transfer
through the Coulomb interaction is likely to be inefficient.

On the other hand, magnetic field in the shocked matter 
may be in equipartition with the random motion energy of protons
which is directly converted from the kinetic energy of a fireball. 
Recall that, in the well-known equation of magnetohydrodynamics, 
the time evolution of magnetic field, $\partial{\bf B}/\partial t$,
is governed by the diffusion
term and the source term, ${\rm rot} [ {\bf u \times B}]$,
where $\bf u$ is the velocity field of fluid. 
In afterglow shocks there will be turbulent motion of ${\bf u}^2
\sim 1$ and a coherent length scale will be smaller than
the shell thickness of the shocked matter measured in the
shell frame, $\sim r/\gamma$. This suggests that $\left|
{\rm rot}[{\bf u \times
B}]\right| \gtilde (\gamma/r)B$ and hence the growth time scale of magnetic
field may be $\ltilde r/\gamma$. Since the expansion time is also
$\sim r/\gamma$ in the shell frame, it is possible that equipartition
between magnetic field and protons is achieved while electrons 
carry much smaller energy. Although the above argument is quite rough
and some unknown processes in relativistic matter
may allow electrons to be in equipartition with protons,
it seems rather reasonable to consider the case of $\xi_B \sim 1$ and
$\xi_e \sim (m_e/m_p) \ll 1$. In order to investigate such an energetic
model of GRBs, we use $E = 10^{52} E_{52}$ erg and $b = 200 b_{200}$ 
as typical values,
with which $E$ is about 2000 times larger than the energy emitted in
GRB photons, i.e., $\sim 10^{51} b^{-1}$ erg when $z_{\max} \sim 1$.

\section{Proton Synchrotron in Afterglow Shock}
If the energy transfer from protons into electrons is inefficient but
magnetic field is nearly in equipartition, the synchrotron radiation of
protons becomes relatively important. The energy density of shocked matter
is given by $4\gamma^2 n m_p$ in the shell frame
and magnetic field can be written as
$B = (32\pi \xi_B \gamma^2 n m_p)^{1/2}$.
It has been considered that protons may be accelerated up to $\sim 10^{20}$
eV in GRBs because the physical quantities of GRBs
allow acceleration of protons
to such high energies and observed flux of
highest energy cosmic rays is consistent with the GRB occurrence rate
provided that such protons carry roughly the same amount
of energy with GRBs (Waxman 1995; Vietri 1995). We assume that
the shock acceleration time is given by $\eta r_L$, where $r_L
= m_p \gamma_p/(eB)$ is
the Larmor radius, $\gamma_p$ is the proton Lorentz factor in
the shell frame, and $\eta$ is a parameter of order unity. 
The maximum energy obtained in the external shock 
is given by the equation $\eta r_L = r/\gamma$, and we find
$4.21 \times 10^{21} \eta^{-1} 
\xi_B^{1/2} n_1^{1/6} \gamma_{100}^{1/3} b_{200}^{1/3} E_{52}^{1/3}$ eV
in the observer's frame, where $\gamma_{100} = \gamma_0/100$
and $\gamma_0$ is the initial fireball Lorentz factor.
On the other hand, 
the maximum energy is also constrained by synchrotron cooling.
The cooling time at the shell frame
is $t_{\rm syn} = 3 m_p^3 / (4 \sigma_T m_e^2 U_{\rm mag}
\gamma_p)$, where $U_{\rm mag}$ is the energy density of magnetic field.
From the equation $\eta r_L = t_{\rm syn}$ we then find
the maximum energy in the observer's frame as
$3.27 \times 10^{21} \eta^{-1/2} \xi_B^{-1/4}
n_1^{-1/4} \gamma_{100}^{1/2}$ eV, which does not depend on the total 
fireball energy
of GRBs. Therefore protons may by accelerated up to $10^{21-22}$ eV
for $\gamma_0$ = 100--1000, 
which is about one order of magnitude greater than the estimate
obtained with $bE \sim 10^{51}$ erg (Waxman 1995; Vietri 1995). 

The protons accelerated up to $\sim 10^{21}$ eV 
will radiate their energy by synchrotron radiation 
in magnetic fields of afterglow
shock. The synchrotron photon energy in the observer's frame is
given by $\varepsilon_{\gamma} =
\gamma \gamma_p^2 e B / m_p = 2.8 \varepsilon_{p, 21}^2 
\xi_B^{1/2} n_1^{1/2} \ \rm TeV \ ,$
where $\varepsilon_{p, 21} = m_p \gamma \gamma_p /(10^{21}$eV) 
is proton energy in the observer's frame. Therefore synchrotron photon
energy is related to proton energy independently of time and 
will extend to $\gtilde$ TeV. Next let us check whether
such protons can be confined in the afterglow shock. The ratio of
Larmor radius of protons to the restframe shell thickness $r/\gamma$
is found as $
r_{L}/(r/\gamma) = 0.26 \ \xi_B^{-3/4} n_1^{-3/8} b_{200}^{-3/8}
E_{52}^{-3/8} \varepsilon_{\rm \gamma, TeV}^{1/2} t_{\rm day}^{1/8}$,
where $\varepsilon_\gamma = \varepsilon_{\gamma, \rm TeV}$ TeV and
the observation time is $t_{\rm day} \ {\rm day}
= r/(2\gamma^2)$ or $\gamma = 15 \ t_{\rm day}^{-3/8} b_{200}^{1/8} 
E_{52}^{1/8} n_1^{-1/8}$. The transverse width of beamed shell is larger
than the shell thickness unless the beaming factor is extremely large
as $b > 7.21 \times 10^3 t_{\rm day}^{-1} E_{52}^{1/3} n_1^{-1/3}$.
Therefore the protons which correspond to synchrotron photons of
$\ltilde$ TeV can be trapped within the magnetic field of afterglow
shock on a time scale of $\sim$ day.
The cooling time observed on the Earth 
is related to the restframe cooling time as $t_{\rm syn, obs} =
t_{\rm syn}/(2\gamma)$, and we find 
$t_{\rm syn, obs} =
1.3 \ \xi_B^{-3/4} n_1^{-1/2} \varepsilon_{\gamma, \rm TeV}^{-1/2}
t_{\rm day}^{3/4} b_{200}^{-1/4} E_{52}^{-1/4} \ \rm day$.
If the spectrum of accelerated protons is that of the standard
shock acceleration theory, i.e., $dN_p/d\gamma_p \propto \gamma_p^{-\alpha}$
with $\alpha \sim 2$, luminosity of synchrotron radiation per decade
of photon energy, $L(\varepsilon_\gamma) \equiv \varepsilon_\gamma
dL/d\varepsilon_\gamma$, is proportional to $\varepsilon_\gamma^{\beta}$
with $\beta = (3 - \alpha)/2 \sim 0.5$. Hence
the synchrotron emissivity becomes maximum at the cut-off energy and
the above results suggest the following picture: protons
accelerated to $\sim 10^{20-21}$ eV will be trapped in afterglow shock and
radiate most of their energy 
in the TeV range within a time scale of a few days.
Since the total energy of protons could be 
comparable to the total kinetic energy
of a fireball while electrons carry much smaller energy, the energy
radiated by proton synchrotron around TeV range would be much larger
than the energy released as GRB photons.

Now we proceed to estimate of the luminosity of proton synchrotron
radiation. Suppose that accelerated protons in shocked matter
have a power-law spectrum ($\alpha = 2$)
in the range $\gamma_0^2 \leq \tilde\gamma_p
\leq \gamma_u \sim 10^{12}$ and total kinetic energy carried by them
in the observer's frame is $\xi_p E$:
\begin{equation}
\frac{dN_p}{d\tilde\gamma_p} = \frac{\xi_p E}{m_p \ln(\gamma_u/\gamma_0^2)}
\tilde\gamma_p^{-2} \ ,
\end{equation}
where $\tilde\gamma_p = \gamma \gamma_p$
is the proton Lorentz factor at the observer's
frame and $\xi_p \sim 1$ if the accelerated protons are in equipartition.
The observed luminosity is given by
\begin{equation}
L(\varepsilon_\gamma) \equiv 
\varepsilon_\gamma \frac{dL}{d\varepsilon_\gamma} =
2 \gamma^2 \varepsilon_{\gamma}
\frac{d\gamma_p}{d\varepsilon_{\gamma}}
\frac{dN_p}{d\gamma_p} \ j_{\rm syn, p} \ ,
\end{equation}
where $j_{\rm syn, p} = 4 \sigma_T m_e^2 U_{\rm mag} \gamma_p^2
/ (3 m_p^2)$ is the synchrotron
energy loss rate of a proton in the shell frame. 
After some calculations we find 
\begin{equation}
L(\varepsilon_\gamma) =
2.5 \times 10^{45} \ \xi_p \xi_B^{3/4} E_{52}^{5/4} b_{200}^{1/4}
n_1^{1/2} t_{\rm day}^{-3/4} \varepsilon_{\gamma, \rm TeV}^{1/2}
\ \rm erg \ s^{-1} \ ,
\label{eq:luminosity}
\end{equation}
where we have assumed $(\gamma_0, \gamma_u) = (10^2, 10^{12})$.
If we observe this emission from a distance of $d = 3000 d_{3}$ Mpc
($z \sim 1$), then the observed flux above 1 TeV is $5.9 \times 10^{-10} 
\ \xi_p \xi_B^{3/4} E_{52}^{5/4} b_{200}^{5/4}
n_1^{1/2} t_{\rm day}^{-3/4} 
d_3^{-2} \rm \ photons \ cm^{-2} \ sec^{-1}$. This flux is further 
attenuated by the $e^\pm$ creation with intergalactic infrared photon field.
The current estimate of
the optical depth for this intergalactic absorption
is still highly uncertain, but a typical value for
TeV gamma-rays is $\tau \sim 10$ ($e^\tau = 2.2 \times 10^4$)
for $z \sim 1$ (Salamon \& Stecker 1998). The amount of infrared
background is related to the amount of stars in the universe and this
is uncertain by a factor of about 2. A factor of 2 reduction of the estimate
of $\tau$ results in the attenuation of $e^\tau \sim 150$.
Therefore the above flux would be attenuated 
by a factor of at least 100, and the attenuated flux is 
consistent with the upper limits set by the Whipple telescope (Connaughton
et al. 1997)
for some GRBs, which are about $10^{-10}$--$10^{-9} \ \rm cm^{-2} sec^{-1}$ 
depending on the source position in the field of view ($\sim 3^\circ$).
However, if a burst location is determined as well as some of recent GRBs
for which afterglows are detected, and observation is made for a time scale of
day, a flux of $\sim 10^{-12} \ \rm cm^{-2} sec^{-1}$ is detectable
by currently working 
ground-based air \v{C}erenkov telescopes (see, e.g., Kifune 1996 for
a general review),
and hence the TeV photons from GRBs
at $z \sim 1$ are marginally detectable. 
Detectability increases rapidly with decreasing distance because of
the decrease of optical depth as well as increase of 
the original (unattenuated) flux, and TeV gamma-rays from $z \sim 0.5$
would be easily detectable. 

The above estimate is based on the relatively
small distance scale of GRBs, $z_{\max} \sim 1$, but
larger distance scales are also suggested by cosmic evolution of
star formation rate
(Totani 1997, 1998; Sahu et al. 1997; Wijers et al. 1998) or by the recently
detected host galaxy for GRB971214 (Kulkarni et al. 1998). 
The original TeV flux expected on the Earth 
before absorbed in intergalactic fields is
almost insensitive to the unknown distance scale of GRBs, because
we have just scaled the total energy ($E$) from the observed energy
emitted as GRB photons.
The increase of the intergalactic optical depth with $z$ beyond
$z \sim 1$ is also rather
slow compared to $z<1$ (Salamon \& Stecker 1998), and hence
the detectability of TeV gamma-rays is not so sensitive to
the GRB distance scale, if $z_{\max} \gtilde 1$. 
More precise estimate of detectability requires
better determination of infrared background, and in other words, 
discovery of the TeV afterglow would give important
infomation for the intergalactic infrared photon field.

\section{Detectability of GeV Photons}
GRB940217 has the third-largest energy fluence in the 4B BATSE catalog
(Paciesas et al. 1997),
and this GRB is famous for the detection of high energy photons
by the EGRET detector with very long duration (Hurley et al. 1994). The EGRET
detected high energy photons ranging from 36 MeV to 18 GeV 
during $\sim$ 5000 seconds.
We show that this EGRET photons are well explained by the proton synchrotron
of the model. The observer's time when the external shock phase begins, 
$t_d$, is given by $r_d/(2 \gamma_0^2)$, where the deceleration radius is
$r_d = (17 b E / 16\pi n m_p \gamma_0^2)^{1/3}$. This deceleration time 
sensitively depends on $\gamma_0$, and it can be as short as $t_d = 
1.3 \ b_{200}^{1/3} E_{52}^{1/3} n_1^{-1/3} \gamma_{1000}^{-8/3}$ sec
when $\gamma_0 \sim$ 1000. Therefore the EGRET photons can be considered as
the external shock origin and our model is applicable, 
although other explanations by internal shocks
may also be possible.
The observed photon spectrum (Fig. 3 of Hurley et al. 1994)
seems consistent with
the standard spectrum of synchrotron radiation, $dn/d\varepsilon_\gamma
\propto \varepsilon_\gamma^{-3/2}$, 
and by fitting the data with this photon index we find that
the differential photon flux $dn/d\varepsilon_\gamma$ is $\sim 2 \times
10^{-11}$ and $2 \times 10^{-12} \varepsilon_{\gamma, \rm GeV}^{-3/2} \
\rm photons \ cm^{-2} \ sec^{-1} \ keV^{-1}$ for the first 180 sec
and delayed photons (180--5400 sec), respectively.
This time evolution is consistent with the $t^{-3/4}$ 
profile of Eq. (\ref{eq:luminosity}). 
If we assume that the fluence of this GRB in the BATSE range,
$6.6 \times 10^{-4} \ \rm erg \ cm^{-2}$, is 1/2000 of the
total energy $E$, the distance to this GRB is $d = 113 \ b_{200}^{1/2}
E_{52}^{1/2}$ Mpc and hence the differential photon flux obtained
from Eq. (\ref{eq:luminosity})
is $5.6 \times 10^{-11} \xi_p \xi_B^{3/4} E_{52}^{1/4} b_{200}^{1/4}
n_1^{1/2} t_{5}^{-3/4} \varepsilon_{\rm \gamma, GeV}^{-3/2} \
\rm photons \ cm^{-2} \ sec^{-1} \ keV^{-1}$,
where $t_5 = t_{\rm obs}/(5000$ sec). 
This photon flux is consistent with the observation
if the energy conversion into accelerated protons and
magnetic field is near the equipartition: $\xi_p \sim \xi_B^{3/4} \sim 0.2
n_1^{-1/4}$.
Therefore the delayed GeV photons from GRB940217 are naturally explained
by our model. On the other hand, there exist some 
GRBs which are as bright as the GRB940217 but not accompanied
by such long-duration GeV photons. In such GRBs, the onset of external shock
phase might be very long after the GRBs and/or the density of interstellar
matter is quite low. 
In fact, if the progenitor of a GRB is a massive star,
the intensive stellar wind prior to the death of the star could have
swept up the interstellar medium near the star. In this case the TeV or GeV
luminosity which is proportional to $n^{1/2}$ could be very small and
significantly delayed compared to GRBs.

Because of the $e^{\pm}$ pair-creation in intergalactic field,
more than 99 \% of TeV gamma-rays from $z \sim 1$
must disappear before reaching the
Earth. The created $e^{\pm}$ pairs, whose energy is about $\sim$ TeV,
lose their energy by the inverse-Compton (IC) scattering
of the cosmic microwave background photons and typical energy of the secondary 
IC photons is $ \varepsilon_{2} \sim 0.6
\varepsilon_{\gamma, \rm TeV}^2$ GeV, which is in the
detectable range of the EGRET. 
The expected time delay of these secondary photons
is $\sim d/(2 c \gamma_{\rm pair}^2) =
1.8 \ d_3 \gamma_6^{-2}$ day (Cheng \& Cheng 1996), 
where $\gamma_{\rm pair} = 10^6 \gamma_6$ is the Lorentz factor of
created pair. If the attenuation of TeV gamma-rays is significant
($e^\tau \gg 1$), almost all energy originally emitted in TeV range should be
converted into GeV range, which is much larger than the original
energy emitted in GeV photons by proton synchrotron. 
This effect becomes significant with increasing optical
depth for TeV photons and compensate the decrease of flux due to the increase 
of distance, and hence we might be able to detect delayed GeV photons for 
rather distant GRBs. In the limit of $e^{\tau} \gg 1$ and neglecting
the time delay due to propagation in the intergalactic field,
we have estimated the differential photon flux as $
1.7 \times 10^{-7} \xi_p \xi_B^{3/4} E_{52}^{5/4} b_{200}^{5/4}
n_1^{1/2} t_{\rm day}^{-3/4} d_3^{-2} \varepsilon_{2, \rm GeV}^{-7/4} \ \rm
photons \ GeV^{-1} \ cm^{-2} \ s^{-1}$. This estimate may be further 
reduced by the delay of $\sim$ a few days, but not so far from 
the EGRET sensitivity. 
Delayed GeV emission on a time scale of a few days from GRBs at 
cosmological distances may be detectable by the EGRET, or is likely
to be detected by the future GLAST experiment. Future ground-based telescopes
with reduced threshold
energy down to tenth of GeV range,
e.g., the VERITAS project (Weekes et al. 1998) will also be useful 
for search of the secondary GeV photons.

\section{Discussion}
The typical Lorentz factor of electrons in afterglow shock is
$\gamma_e = \xi_e (m_p/m_e)\gamma$ and we have considered the
case of $\xi_e \sim 1/2000$. Then the observed synchrotron photon
energy of a electron with a Lorentz factor $\gamma_e$
is $\gamma \gamma_e^2 e B / m_e =
2.5 \times 10^{-4} \ (2000\xi_e)^2
\xi_B^{1/2} b_{300}^{1/2} E_{52}^{1/2} t_{\rm day}^{-3/2}$ eV, 
which is the radio band, and it seems to contradict the observations
of X-ray or optical afterglows. However acceleration of electrons
and/or IC scattering of synchrotron photons can raise
the photon energy. Furthermore, $\xi_e$ may also increase with
time in afterglow. In fact, $\tau_{ep}/(r/\gamma)
= 870 b_{200}^{-1/3} E_{52}^{-1/3} n_1^{-2/3} \gamma^{8/3}$ decreases as
$\propto \gamma^{8/3} \propto t^{-1}$, and it is possible that
energy transfer from protons into electrons becomes
efficient gradually with the expansion of afterglow shock. 
Note that there are considerable variations for the behavior of
afterglows observed in X, optical, and radio bands (e.g., Groot et al. 1998). 
Efficiency of energy transfer into electrons and
its time evolution could have large variations among GRBs and it may be
one of the origins of the complicated behavior of
GRB afterglows. 

It should be noted that the proton synchrotron
emission extends to X-ray, optical, and radio bands with the
standard synchrotron spectrum of $dL/d\varepsilon_\gamma \propto
\varepsilon^{(1-\alpha)/2}$. If $\alpha = 2$, luminosity per decade
of photon energy at $\varepsilon_\gamma$ = 1 keV
is $\sqrt{10^9}$ times smaller than
that in the TeV range, and the differential
flux observed from a distance of 3000 Mpc
is $\sim 1.5 \times 10^{-14} \xi_p \xi_B^{3/4} E_{52}^{5/4} b_{200}^{5/4}
n_1^{1/2} t_{\rm day}^{-3/4} \varepsilon_{\gamma, \rm keV}^{-1/2}
d_3^{-2} \ \rm erg \ cm^{-2} \ sec^{-1} \ keV^{-1}$. This flux is comparable
to the observed flux of X-ray afterglows for distant bursts such as
GRB970402 or GRB970508 (Piro et al. 1997a,b).
Therefore the proton synchrotron radiation could
contribute to the X-ray afterglows, although it is rather difficult to
detect in the 
optical or lower energy bands due to the hardness of the spectrum.
Note that optical afterglows are associated
only to a small fraction of GRBs for which
X-ray afterglows are detected. Therefore it can be speculated that
proton synchrotron was dominant in such GRBs. The complicated
behavior of afterglows may be a consequence of complicated
mixture of proton synchrotron and electron synchrotron or
inverse Compton scattering.

We finally note that energy emitted as $10^{20-21}$ eV protons must be
roughly the same with that emitted as GRB photons, if
the GRB is the origin of ultra high energy cosmic rays (UHECRs)
observed on the Earth (Waxman 1995; Vietri 1995).
On the other hand, in our model, energy distributed to such protons is
much greater (at least by a factor of $\sim 100$) than GRB photons.
However, as we have shown, such protons are likely trapped in 
afterglow shock and lose their energy by synchrotron radiation. 
If the escape fraction of protons just cancels the overproduction of 
$10^{20}$ eV protons, GRBs could still be the origin of UHECRs. 
If the escape fraction is further smaller, then the UHECRs
must be explained by other sources. The Larmor radius
becomes larger with increasing proton energy, 
and the escape fraction may increase with proton energy.
This suggests a possibility that the spectrum of UHECRs becomes 
significantly harder above $10^{20}$ eV, which should be tested
by future experiments.

The author would like to thank S. Sasaki and an anonymous referee
for useful comments.
He has been supported by the Research Fellowships of the Japan
Society for the Promotion of Science for Young Scientists, and
the Grant-in-Aid for the
Scientific Research Fund (No. 3730) of the Ministry of Education, Science,
and Culture of Japan.

\end{document}